\documentclass[conference]{IEEEtran}

\IEEEoverridecommandlockouts
\usepackage[letterpaper, left=1in, right=1in, bottom=1in, top=0.75in]{geometry}

\usepackage{ucs}
\usepackage[utf8x]{inputenc}
\usepackage[cmex10]{amsmath}
\usepackage{cite, amsfonts, amssymb, amsthm, bm, bbm, graphicx, relsize, multirow, booktabs, tikz,subfigure,soul}
\usepackage[american]{babel}
\usepackage[T1]{fontenc}
\usepackage{algorithmic, algorithm,color}
\usepackage[multiple]{footmisc}
\usepackage{glossaries}
\newtheorem{proposition}{Proposition}
\newtheorem{lemma}{Lemma}

\theoremstyle{definition}

\theoremstyle{remark}

\newcommand{\bL}{\mathbf{L}}
\newcommand{\bA}{\mathbf{A}}
\newcommand{\bbeta}{\boldsymbol{\eta}}
\newcommand{\bx}{\boldsymbol{x}}

\newcommand{\ds}{\displaystyle}

\makeglossaries

\newacronym{see}{SEE}{secrecy energy efficiency}
\newacronym{miso}{MISO}{multiple input single output}
\newacronym{miso-se}{MISO-SE}{multiple input single output single-antenna eavesdropper}
\newacronym{lmmse}{LMMSE}{linear minimum mean square error}
\newacronym{d2d}{D2D}{device-to-device}
\newacronym{p2p}{P2P}{point-to-point}
\newacronym{mac}{MAC}{multiple-access channel}
\newacronym{bc}{BC}{broadcast channel}
\newacronym{ic}{IC}{interference channel}
\newacronym{imac}{IMAC}{interference multiple access channel}
\newacronym{ibc}{IBC}{interference broadcast channel}
\newacronym{mimo}{MIMO}{multiple-input multiple-output}
\newacronym{mimo-me}{MIMO-ME}{multiple input multiple output multiple-antenna eavesdropper}
\newacronym{siso}{SISO}{single-input single-output}
\newacronym{sc}{SC}{single-carrier}
\newacronym{mc}{MC}{multi-carrier}
\newacronym{ofdma}{OFDMA}{orthogonal frequency division multiple access}
\newacronym{af}{AF}{amplify-and-forward}
\newacronym{df}{DF}{decode-and-forward}
\newacronym{cf}{CF}{compress-and-forward}
\newacronym{mwrc}{MWRC}{multi-way relay channel}
\newacronym{pde}{PDE}{partial data exchange}
\newacronym{fde}{FDE}{full data exchange}
\newacronym{iid}{i.i.d.\@}{independent and identically distributed}
\newacronym{awgn}{AWGN}{additive white Gaussian noise}
\newacronym{awg}{AWG}{additive white Gaussian}
\newacronym{sic}{SIC}{successive interference cancellation}
\newacronym{dpc}{DPC}{dirty paper coding}
\newacronym{snr}{SNR}{signal-to-noise ratio}
\newacronym{sinr}{SINR}{signal to interference plus noise ratio}
\newacronym{ber}{BER}{bit error rate}
\newacronym{zf}{ZF}{zero-forcing}
\newacronym{mmse}{MMSE}{minimum mean square error}
\newacronym{sud}{SUD}{single user decoding}
\newacronym{dof}{DoF}{degrees of freedom}
\newacronym{gdof}{GDoF}{generalized degrees of freedom}
\newacronym{nnc}{NNC}{noisy network coding}
\newacronym{dmn}{DMN}{discrete memoryless network}
\newacronym{csi}{CSI}{channel state information}
\newacronym{ee}{EE}{energy efficiency}
\newacronym{ian}{IAN}{treating interference as noise}
\newacronym{snd}{SND}{simultaneous non-unique decoding}
\newacronym{brd}{BRD}{best response dynamics}
\newacronym{br}{BR}{best response}
\newacronym{ne}{NE}{Nash equilibrium}
\newacronym{lhs}{LHS}{left-hand side}
\newacronym{rhs}{RHS}{right-hand side}
\newacronym{gee}{GEE}{global energy efficiency}
\newacronym{wsee}{WSEE}{weighted sum energy efficiency}
\newacronym{wpee}{WPEE}{weighted product energy efficiency}
\newacronym{wmee}{WMEE}{weighted minimum energy efficiency}
\newacronym{kkt}{KKT}{Karush Kuhn Tucker}
\newacronym{pc}{PC}{pseudo-concave}
\newacronym{qc}{QC}{quasi-concave}
\newacronym{ql}{QL}{quasi-linear}
\newacronym{pl}{PL}{pseudo-linear}
\newacronym{spc}{SPC}{strictly pseudo-concave}
\newacronym{sqc}{SQC}{strictly quasi-concave}
\newacronym{lfp}{LFP}{linear fractional problem}
\newacronym{clfp}{CLFP}{concave-linear fractional problem}
\newacronym{ccfp}{CCFP}{concave-convex fractional problem}
\newacronym{mmfp}{MMFP}{max-min fractional problem}
\newacronym{sorp}{SoRP}{sum-of-ratios problem}
\newacronym{porp}{PoRP}{product-of-ratios problem}
\newacronym{qos}{QoS}{quality of service}
\newacronym{evd}{EVD}{eigenvalue decomposition}
\newacronym{svd}{SVD}{singular value decomposition}
\newacronym{skee}{SKEE}{Secret-key energy efficiency}
\newacronym{an}{AN}{artificial noise}
\usepackage{cite}

\ifCLASSINFOpdf
\else
\fi
\usepackage[cmex10]{amsmath}

\usepackage{algorithmic}

\IEEEoverridecommandlockouts \IEEEpubid{\makebox[\columnwidth]{ 978-1-5386-3531-5/17/\$31.00~\copyright~2017 IEEE \hfill} \hspace{\columnsep}\makebox[\columnwidth]{ }}

\begin{document}
\title{Downlink Power Control in User-Centric and Cell-Free Massive MIMO Wireless Networks}

\author{\IEEEauthorblockN{Stefano Buzzi  and Alessio Zappone}
\IEEEauthorblockA{Department of Electrical and Information Engineering\\
University of Cassino and Lazio Meridionale, Cassino, Italy}
\thanks{The work of A. Zappone has been supported by the PRASG project, funded by the University of Cassino and Southern Lazio.}}
%\and
%\IEEEauthorblockN{Homer Simpson}
%\IEEEauthorblockA{Twentieth Century Fox\\
%\and
%\IEEEauthorblockN{James Kirk\\ and Montgomery Scott}
%\IEEEauthorblockA{Starfleet Academy\\
%San Francisco, California 96678-2391\\


%Telephone: (800) 555--1212, Fax: (888) 555--1212}



\maketitle

\begin{abstract}
Recently, the so-called cell-free  Massive MIMO architecture has been introduced, wherein a very large number of distributed access points (APs) simultaneously and jointly serve a much smaller number of
mobile stations (MSs). A variant of the cell-free technique is the user-centric approach, wherein each AP just decodes the MSs that it receives with the largest power. This paper considers both the cell-free and user-centric approaches, and, using an interplay of sequential optimization and alternating optimization, derives downlink power-control algorithms aimed at maximizing either the minimum users' SINR (to ensure fairness), or the system sum-rate. 
Numerical results show the effectiveness of the proposed algorithms, as well as that the user-centric approach generally outperforms the CF one.
\end{abstract}


% \fi
\IEEEpeerreviewmaketitle

\section{Introduction}
Cell-free (CF) massive MIMO is a recent system concept \cite{ngo2015cell,Ngo_CellFree2016} addressing the cell-edge problem: a very large number of distributed APs (connected via a backhaul network to a central CPU) simultaneously and jointly serve a much smaller number of
MSs; each AP uses local channel estimates obtained from received uplink pilots and applies channel inversion beamforming to transmit data to the users.  
Papers  \cite{ngo2015cell,Ngo_CellFree2016} show that the CF approach provides better performance than a small-cell system in terms of 95$\%$-likely per-user throughput. Additionally, the paper \cite{nayebi2015cell} has analyzed the performance improvements granted by the use of a zero-forcing precoder in the downlink. 
In \cite{WSA2017_cell_free} the CF approach is extended to the case in which multiple antennas are employed and propose the use of a channel-independent beamformer at the MSs, so that channel estimation is performed only in the uplink at the APs, and channel reciprocity induced by time division duplex is exploited at the APs to implement coherent signal reception at the MSs. 
The paper \cite{WSA2017_cell_free} also introduces a user-centric (UC) approach wherein each AP chooses to serve only a pre-determined number of MSs. The UC approach is shown to provide larger achievable rates (with respect to the CF approach) to the vast majority of the users, and, also works with reduced backhaul overhead; a comparison between the CF and the UC approaches in a single-antenna setting is provided in \cite{WCL2017buzzidandrea}. While the papers \cite{WSA2017_cell_free,WCL2017buzzidandrea} consider a simple uniform power control allocation on the downlink, in this paper we provide power control algorithms aimed at maximizing either the minimum rate across the users -- to ensure fairness -- or the system sum rate. The considered optimization problems are non-convex and involve a very large number of variables (equal to the product of the number of APs times the number of MSs); 
to address these challenges, the frameworks of sequential optimization and alternating optimization are combined in a suitable way. Our results show that the proposed power control algorithms are effective, as well as that the UC approach generally outperforms the CF one. 

%The remainder of this paper is organized as follows. Next Section briefly describes the considered system model (see \cite{WSA2017_cell_free} for additional details). Section III is devoted to the illustration of the downlink power control schemes for both the CF and UC approaches, while Section IV contains the numerical results. Finally, concluding remarks are given in Section V.

\section{System model}
We consider an area with $K$ MSs and $M$ APs. MSs and APs are randomly located. 
The $M$ APs are connected by means of a backhaul network to a central processing unit (CPU) wherein data-decoding is performed. The TDD protocol is used and 
the channel coherence interval is divided into three phases: (a) uplink channel estimation, (b) downlink data transmission, and (c) uplink data transmission. In phase (a) the MSs send pilot data in order to enable channel estimation at the APs. 
In phase (b) the APs use channel estimates to perform channel-matched beamforming and send data symbols on the downlink; while in the CF architecture the APs send data to all the MSs in the system, in the UC approach each AP sends data only to a subset of the MSs in the system. Phase (c) is not described since the focus of this paper is only on the downlink. 

\subsection{Channel model}
We denote by the $(N_{\rm AP} \times N_{\rm MS})$-dimensional matrix $\mathbf{G}_{k,m}$ the channel between the $k$-th MS and the $m$-th AP. We have
\begin{equation}
\mathbf{G}_{k,m}=\beta_{k,m}^{1/2} \mathbf{H}_{k,m} \; ,
\label{channel_model}
\end{equation}
with $\beta_{k,m}$ a scalar coefficient modeling the channel shadowing effects and 
$\mathbf{H}_{k,m}$ an $(N_{\rm AP} \times N_{\rm MS})$-dimensional matrix whose entries are i.i.d ${\cal CN}(0,1)$ RVs. Path loss and shadow fading are modeled by the large scale coefficient $\beta_{k,m}$ in \eqref{channel_model}, according to the formula \cite{Ngo_CellFree2016}
\begin{equation}
\beta_{k,m}= 10^{\frac{\text{PL}_{k,m}}{10}} 10^{\frac{\sigma_{\rm sh}z_{k,m}}{10}},
\end{equation}
where $\text{PL}_{k,m}$ represents the path loss (expressed in dB) from the $k$-th MS to the $m$-th AP, and $10^{\frac{\sigma_{\rm sh}z_{k,m}}{10}}$ represents the shadow fading with standard deviation $\sigma_{\rm sh}$, while $z_{k,m}$ will be specified later.
For the path loss we use the following three slope path loss model \cite{tang2001mobile}:
\begin{equation}
\text{PL}_{k,m}=\left\lbrace 
\begin{array}{lll}
-L-35 \log_{10}\left(d_{k,m}\right), & & \text{if} \; d_{k,m}>d_1 \\
-L - 10 \log_{10}\left(d_1^{1.5} d_{k,m}^{2}\right), & & \text{if} \; d_0< d_{k,m}\leq d_1 \\
-L - 10 \log_{10}\left(d_1^{1.5} d_{0}^{2}\right), & & \text{if} \; d_{k,m}<d_0
\end{array} \right. ,
\label{path_loss}
\end{equation} 
where $d_{k,m}$ denotes the distance between the $m$-th AP and the $k$-th user, $L$ is
\begin{equation}
\begin{array}{lll}
L=& 46.3+33.9\log_{10}\left(f\right)-13.82\log_{10}\left(h_{\rm AP}\right)- 
\\ & \left[1.11\log_{10}\left(f\right)-0.7\right]h_{\rm MS} +1.56\log_{10}\left(f\right)-0.8,
\end{array}
\end{equation}
$f$ is the carrier frequency in MHz, $h_{\rm AP}$ and $h_{\rm MS}$ denote the AP and MS antenna heights, respectively.
In real-world scenarios, transmitters and receivers that are in close vicinity of each other may be surrounded by common obstacles, and hence, the shadow fading RVs are correlated; for the shadow fading coefficient we thus use a model with two components \cite{wang2008joint}
\begin{equation}
z_{k,m}=\sqrt{\delta}a_m+\sqrt{1-\delta}b_k, \; \; m=1, \ldots, M, \; k=1,\ldots,K,
\label{shadowing}
\end{equation}
where $a_m \sim \mathcal{N}(0,1)$ and $b_k \sim \mathcal{N}(0,1)$ are independent RVs, and $\delta, \; 0\leq \delta \leq 1$ is a parameter.
The covariance functions of $a_m$ and $b_k$ are given by:
\begin{equation}
E\left[a_m a_{m'}\right]=2^{-\frac{d_{\rm{AP}(m,m')}}{d_{\rm decorr}}} \; \; \; \; 
E\left[b_k b_{k'}\right]=2^{-\frac{d_{\rm{MS}(k,k')}}{d_{\rm decorr}}},
\label{shadowing_corr}
\end{equation}
where $d_{\rm{AP}(m,m')}$ is the geographical distance between the $m$-th and $m'$-th APs and  $d_{\rm{MS}(k,k')}$ is the geographical distance between the $k$-th and the $k'$-th MSs. The parameter $d_{\rm decorr}$  is a decorrelation distance which depends on the environment, typically this value is in the range 20-200 m.

\subsection{Uplink training}
During this phase the MSs send uplink training pilots in order to permit channel estimation at the APs.  This phase is the same for both the UC and CF approaches.
We denote by $\tau_c$ the length (in samples) of the channel coherence time, and by $\tau_p$ the length (in samples) of the uplink training phase. Of course we must ensure that $\tau_p < \tau_c$. 
Denote by $\mathbf{\Phi}_k \in {\cal C}^{N_{\rm MS}\times \tau_p}$ the pilot sequence sent by the $k$-th MS, and assume that $\|\mathbf{\Phi}_k\|^2_F=1$;  we also assume that the pilot sequences assigned to each user are mutually orthogonal, so that $\mathbf{\Phi}_k \mathbf{\Phi}_k^H=\mathbf{I}_{N_{\rm MS}}$, while, instead, pilot sequences from different users are non-orthogonal. 
The signal received at the $m$-th AP in the $n$-th signaling time is represented by the following $N_{\rm AP}$-dimensional vector:
\begin{equation}
\mathbf{y}_m(n)= \ds \sum_{k=1}^K \ds \sqrt{p}_k \mathbf{G}_{k,m}\mathbf{\Phi}_k(:,n) + \mathbf{w}_m(n) \; ,
\end{equation}
with $\sqrt{p}_k$ the $k$-th user's transmit power during the training phase.
Collecting all the observable vectors  $\mathbf{y}_m(n)$, for $n=1, \ldots, \tau_p$ into the $(N_{\rm AP} \times \tau_p)$-dimensional matrix $\mathbf{Y}_m$, and assuming 
simple pilot-matched (PM) single-user channel estimation, the estimate,  
$\widehat{\mathbf{G}}_{k,m}$ say, of the channel matrix ${\mathbf{G}}_{k,m}$ is obtained at the $m$-th AP according to $\widehat{\mathbf{G}}_{k,m}=  \ds \frac{1}{\sqrt{p_k}}\mathbf{Y}_m \mathbf{\Phi}_k^H$.

\subsection{Downlink data transmission}
After that each AP has obtained estimates of the channel matrix from all the MSs in the system, the downlink data transmission phase begins. 
The APs treat the channel estimates as the true channels, and channel inversion  beamforming is performed to transmit data to the MSs. 
Denoting by $P_k$ the multiplexing order (i.e., the number of simultaneous data-streams) for user $k$,  and by $\mathbf{x}_k^{\rm DL}(n)$ the $P_k$-dimensional unit-norm vector containing the $k$-th user data symbols to be sent in the $n$-th sample time, and letting $\mathbf{L}_k=\mathbf{I}_{P_k} \otimes \mathbf{1}_{N_{\rm MS}/P_k}$, the downlink precoder at the $m$-th AP for the $k$-th MS is expressed as 
\begin{equation}
\mathbf{Q}^{\rm DL}_{k,m}=\ds \frac{\widehat{\mathbf{G}}_{k,m} \left(\widehat{\mathbf{G}}_{k,m}^H \widehat{\mathbf{G}}_{k,m} \right)^{-1} \mathbf{L}_k}{
\sqrt{\mbox{tr}\left(\mathbf{L}_k^H  \left(\widehat{\mathbf{G}}_{k,m}^H \widehat{\mathbf{G}}_{k,m} \right)^{-1} \mathbf{L}_k 
\right)}} \; .
\label{eq:downlink_precoder}
\end{equation}

\subsubsection{UC massive MIMO architecture}
In the user-centric approach, we assume that each AP communicates only with a fixed number $N$ of MSs.
The generic $m$-th AP, based on the knowledge of the channel estimates $\widehat{\mathbf{G}}_{k,m}$, for $k=1, \ldots, K$, sorts them in descending order according to the Frobenius norm, and selects the first $N$. 
We denote by $\mathcal{K}(m)$ the set of MSs served by the $m$-th AP. 
%and it is
%\begin{equation}
%\mathcal{K}(m)=\{ k : \|\widehat{\mathbf{G}}_{k,m}\|_F \geq \mathbf{\bar{G}}_{m} \}
%\end{equation}
Given the sets $\mathcal{K}(m)$, for all $m=1, \ldots, M$,  we can define the set $\mathcal{M}(k)$  of the APs that communicate with the $k$-th user:
\begin{equation}
\mathcal{M}(k)=\{ m: \,  k \in \mathcal{K}(m) \}
\end{equation}
So, in this case, the signal transmitted by the $m$-th AP in the $n$-th interval is the following $N_{\rm AP}$-dimensional vector
\begin{equation}
\mathbf{s}_m^{\rm uc}(n)=\ds \sum_{k\in{\cal K}(m)} \ds \sqrt{\eta_{k,m}^{\rm DL,uc}} \mathbf{Q}^{\rm DL}_{k,m} \mathbf{x}_k^{\rm DL}(n) \; ,
\end{equation}
with $\eta_{k,m}^{\rm DL,uc}$ a scalar coefficient representing the power transmitted by the $m$-th AP to the $k$-th MS. 
The generic $k$-th MS receives signal contributions from all the APs and so the observable vector is expressed as
\begin{equation}
\begin{array}{llll}
\mathbf{r}_k^{\rm uc}(n)& = & \ds \sum_{m=1}^M
\mathbf{G}_{k,m}^H \mathbf{s}_m^{\rm uc}(n) + \mathbf{z}_k(n) \\
& = &\ds \sum_{m\in{\cal M}(k)} \ds \sqrt{\eta_{k,m}^{\rm DL,uc}} \mathbf{G}_{k,m}^H  \mathbf{Q}^{\rm DL}_{k,m} \mathbf{x}_k^{\rm DL}(n) + \\ & & 
\ds \sum_{j=1, j \neq k}^K \sum_{m\in{\cal M}(j)}   \ds \sqrt{\eta_{j,m}^{\rm DL,uc}}\mathbf{G}_{k,m}^H   \mathbf{Q}^{\rm DL}_{j,m} \mathbf{x}_j^{\rm DL}(n) + \mathbf{z}_k(n)\; .
\end{array}
\label{eq:received_data_MS_UC}
\end{equation}
In \eqref{eq:received_data_MS_UC}, the $N_{\rm MS}$-dimensional vector $\mathbf{z}_k(n)$   represents the thermal noise and out-of-cluster interference at the $k$-th MS, and its entries are modeled
 as i.i.d. ${\cal CN}(0,\sigma^2_z)$ RVs. 
 Based on the observation of the vector $\mathbf{r}_k^{\rm uc}(n)$, a soft estimate of the data symbols 
$\mathbf{x}_k^{\rm DL}(n)$ is obtained at the $k$-th MS as
\begin{equation}
\widehat{\mathbf{x}}_k^{\rm DL,uc}(n)= \mathbf{L}_k^H \mathbf{r}_k^{\rm uc}(n) \; .
\label{Est_DL_uc1}
\end{equation}

\subsubsection{CF massive MIMO architecture} 
In the CF architecture all the APs communicate with all the MSs in the systems, so the CF case can be obtained as a special case of the UC approach letting $N=K$, which leads to ${\cal K}(m)=\{1, \ldots, K\}$, for all $m=1,\ldots, M$ and 
${\cal M}(k)=\{1, \ldots, M\}$, for all $k=1,\ldots, K$.

\section{Downlink power control}
From \eqref{eq:received_data_MS_UC}, we have that the achievable rate for the user $k$ in the user-centric case is written as
\begin{equation}
\mathcal{R}_k=W \log_2 \det \left[\mathbf{I} + \mathbf{R}_k^{-1} \mathbf{A}_{k,k}\mathbf{A}_{k,k}^H \right] \; ,
\label{eq:ASE_expression}
\end{equation}
where 
%\begin{equation}
%\mathbf{A}_{k,k}=\sum_{m\in{\cal M}(k)} \mathbf{L}_k^H \sqrt{\eta_{k,m}^{\rm DL,uc}} \mathbf{G}_{k,m}^H  \widehat{\mathbf{G}}_{k,m} \left(\widehat{\mathbf{G}}_{k,m}^H \widehat{\mathbf{G}}_{k,m} \right)^{-1} \mathbf{L}_k \; ,
%\label{eq:Akk_DL}
%\end{equation}
$
\mathbf{R}_k=\sigma^2_z \mathbf{L}_k^H \mathbf{L}_k + \ds \sum_{j=1, j \neq k}^K \mathbf{A}_{k,j}\mathbf{A}_{k,j}^H \; ,
$
and
\begin{equation}
\mathbf{A}_{k,j}\!=  \mathbf{L}_k^H \!\!\!\! \sum_{m\in{\cal M}(j)} \!\!\ds \sqrt{\eta_{j,m}^{\rm DL,uc}}\mathbf{G}_{k,m}^H   
\widehat{\mathbf{G}}_{j,m} \left(\widehat{\mathbf{G}}_{j,m}^H \widehat{\mathbf{G}}_{j,m} \right)^{-1} \!\!\mathbf{L}_j\; .
\label{eq:Akj_DL}
\end{equation}
In order to obtain the achievable rate expressions for the classical cell-free MIMO case, it suffices to consider in the above expressions ${\cal M}(k)=\{1, 2, \ldots, K\}$, for all $k=1, \ldots, K$.

The rest of this work will be concerned with the optimization of the downlink transmit powers for the maximization of the system sum-rate and minimum users' rate, subject to maximum power constraints.  Mathematically, the sum-rate maximization problem is formulated as the optimization program:
\begin{subequations}\label{Prob:SumRate}
\begin{align}
&\ds\max_{\bbeta}\;\sum_{k=1}^K\mathcal{R}_{k}(\bbeta)\label{Prob:aSumRate}\\
&\;\textrm{s.t.}\; \sum_{k\in{\cal K}_m}\eta_{k,m}\leq P_{max,m}\;,\forall\;m=1,\ldots,M\label{Prob:bSumRate}\\
&\;\;\;\quad\eta_{k,m}\geq 0\;,\forall\;m=1,\ldots,M,\;k=1,\ldots,K\label{Prob:cSumRate}
\end{align}
\end{subequations}
whereas the minimum rate maximization problem is 
\begin{subequations}\label{Prob:MinRate}
\begin{align}
&\ds\max_{\bbeta}\;\min_{1\leq k\leq K}\; \mathcal{R}_{k}(\bbeta)\label{Prob:aMinRate}\\
&\;\textrm{s.t.}\; \sum_{k\in{\cal K}_m}\eta_{k,m}\leq P_{max,m}\;,\forall\;m=1,\ldots,M\label{Prob:bMinRate}\\
&\;\;\;\quad\eta_{k,m}\geq 0\;,\forall\;m=1,\ldots,M,\;k=1,\ldots,K\label{Prob:cMinRate}
\end{align}
\end{subequations}
with $\bbeta$ the $KM\times 1$ vector collecting the transmit powers of all access points. Both problems have non-concave objective functions, which makes their solution challenging. Moreover, even if the problems were concave, the large number of optimization variables, $KM$, would still pose a significant complexity challenge\footnote{Although polynomial, the best known upper-bound for the complexity of generic convex problems scales with the fourth power of the number of variables, while many classes of convex problems admit a cubic complexity \cite{TalNem2001}.}. In order to face these issues, we will resort to the framework of successive lower-bound maximization, recently introduced in\footnote{In \cite{RazaviyaynSIAM} the method is labeled successive upper-bound minimization, since minimization problems are considered.} \cite{RazaviyaynSIAM}, and briefly reviewed next. 

\subsection{Successive lower-bound maximization}\label{Sec:Optimization}
The main idea of the method is to merge the tools of alternating optimization \cite[Section 2.7]{BertsekasNonLinear} and sequential convex programming \cite{SeqCvxProg78}. To elaborate, consider the generic optimization problem 
\begin{align}\label{Prob:GeneralProb}
\ds\max_{\bx\in{\cal X}} f(\bx)\;,
\end{align}
with $f:\mathbb{R}^{n}\to \mathbb{R}$ a differentiable function, and ${\cal X}$ a compact set. As in the alternating optimization method, the successive lower-bound maximization partitions the variable space into $M$ blocks, $\bx=(\bx_1,\ldots,\bx_M)$, which are cyclically optimized one at a time, while keeping the other variable blocks fixed. This effectively decomposes \eqref{Prob:GeneralProb} into $M$ subproblems, with the generic subproblem stated as
\begin{align}\label{Prob:SubProb}
\ds\max_{\bx_{m}} f(\bx_{m},\bx_{-m})\;,
\end{align}
with $\bx_{-m}$ collecting all variable blocks except the $m$-th. It is proved in \cite[Proposition 2.7.1]{BertsekasNonLinear} that iteratively solving \eqref{Prob:SubProb} monotonically improves the value of the objective of \eqref{Prob:GeneralProb}, and converges to a first-order optimal point if the solution of \eqref{Prob:SubProb} is unique for any $m$, and if ${\cal X}={\cal X}_1\times \ldots \times{\cal X}_M$, with $\bx_{m}\in{\cal X}_{m}$ for all $m$. 

Clearly, alternating optimization proves useful when \eqref{Prob:SubProb} can be solved with minor complexity. If this is not the case, the successive lower-bound maximization method proposes to tackle  \eqref{Prob:SubProb} by means of sequential convex programming. This does not guarantee to globally solve \eqref{Prob:SubProb}, but can lead to a computationally feasible algorithm. Moreover, it is guaranteed to preserve the properties of the alternating optimization method \cite{RazaviyaynSIAM}. The idea of sequential optimization is to tackle a difficult maximization problem by solving a sequence of easier maximization problems. To elaborate, let us denote by $g_i(\bx_{m})$ the $i$-th constraint of \eqref{Prob:SubProb},  for $i=1,\ldots,C$. Then, consider a sequence of approximate problems $\{{\cal P}_{\ell}\}_{\ell}$ with objectives $\{f_{\ell}\}_{\ell}$ and constraint functions $\{g_{i,\ell}\}_{i=1}^C$, such that the following three properties are fulfilled, for all $\ell$:
\begin{enumerate}
\item[(\textbf{P1})] $f_{\ell}(\bx_m)\leq f(\bx_m)$, $g_{i,\ell}(\bx_m)\leq g_{i,\ell}(\bx_m)$, for all $i$ and $\bx_m$;
\item[(\textbf{P2})] $f_{\ell}(\bx_m^{(\ell-1)})=f(\bx_m^{(\ell-1)})$, $g_{i,\ell}(\bx_m^{(\ell-1)})=g_{i}(\bx_m^{(\ell-1)})$ with $\bx_m^{(\ell-1)}$ the maximizer of $f_{\ell-1}$;
\item[(\textbf{P3})] $\nabla f_{\ell}(\bx_m^{(\ell-1)})=\nabla f(\bx_m^{(\ell-1)})$, $\nabla g_{i,\ell}(\bx_m^{(\ell-1)})=\nabla g_{i}(\bx_m^{(\ell-1)})$.
\end{enumerate}
In \cite{SeqCvxProg78} (see also \cite{Beck2010,RazaviyaynSIAM}) it is shown that, subject to constraint qualifications, the sequence $\{f(\bx_m^{(\ell)})\}_{\ell}$ of the solutions of the $\ell$-th Problem ${\cal P}_{\ell}$, is monotonically increasing and converges. Moreover, every convergent sequence $\{\bx_m^{(\ell)}\}_{\ell}$ attains a first-order optimal point of the original Problem \eqref{Prob:SubProb}. Thus, the sequential approach enjoys strong optimality properties, fulfilling at the same time the monotonic improvement property for the objective function, and the \gls{kkt} first-order optimality conditions for the original problem. Nevertheless, its applicability hinges on determining suitable lower bounds for the original objective to maximize, which fulfill all three properties \textbf{P1}, \textbf{P2}, \textbf{P3}, while at the same time leading to manageable optimization problems. 

In conclusion, the successive lower-bound maximization method can be formulated as variation of the alternating optimization method, in which each subproblem \eqref{Prob:SubProb} is not globally solved, but instead is tackled by sequential optimization theory. It is proved in \cite{RazaviyaynSIAM} that successive lower-bound maximization has the same optimality properties as the true alternating optimization method, under similar assumptions, even though each subproblem might not be globally solved\footnote{Of course, this holds provided the additional  assumption of the sequential method are fulfilled in each iteration}.
 
\subsection{Sum-rate maximization}\label{Sec:SumRate}
Consider Problem \eqref{Prob:SumRate} and define the variable blocks $\bbeta_{m}$, $m=1,\ldots,M$,  collecting the transmit powers of access point $m$. Then, the sum-rate maximization with respect to the variable block $\bbeta_{m}$ is cast as
\begin{subequations}\label{Prob:SumRateSub}
\begin{align}
&\ds\max_{\bbeta_m}\;\sum_{k=1}^K\mathcal{R}_{k}(\bbeta_m,\bbeta_{-m})\label{Prob:aSumRate}\\
&\;\textrm{s.t.}\; \sum_{k\in{\cal K}_m}\eta_{k,m}\leq P_{max,m}\label{Prob:bSumRateSub}\\
&\;\;\;\quad\eta_{k,m}\geq 0\;,\forall\;k\in {\cal K}_m\label{Prob:cSumRateSub}
\end{align}
\end{subequations}
The complexity of \eqref{Prob:SumRateSub} is significantly lower than that of \eqref{Prob:SumRate}, since only the $M$ transmit powers of access point $m$ are being optimized. Nevertheless, Problem \eqref{Prob:SumRateSub} is still non-convex, which makes its solution difficult. Indeed, defining 
\begin{equation}
\mathbf{A}_{k,j,m}= \mathbf{L}_k^H  \mathbf{G}_{k,m}^H \widehat{\mathbf{G}}_{j,m} \left(\widehat{\mathbf{G}}_{j,m}^H \widehat{\mathbf{G}}_{j,m} \right)^{-1} \mathbf{L}_j\;,
\label{eq:Akjm_DL}
\end{equation}
the $k$-th user's achievable rate can be expressed as 
\begin{align}\label{Eq:kRate}
\mathcal{R}_k(\bbeta)\!\!=\!\!\underbrace{W\!\log_2\!\left|\sigma_z^2\bL_k^H\bL_k\!\!+\!\!\!\sum_{j=1}^K\sum_{m,\ell}\!\!\sqrt{\eta_{j,m}\eta_{j,\ell} }\bA_{k,j,m}\bA_{k,j,\ell}^H\right|}_{g_1(\bbeta)}\\
-\underbrace{W\log_2\left|\sigma_z^2\bL_k^H\bL_k\!+\!\!\!\!\!\sum_{j=1, j \neq k}^K\sum_{m,\ell}\!\!\sqrt{\eta_{j,m}\eta_{j,\ell} }\bA_{k,j,m}\bA_{k,j,\ell}^H\right|}_{g_2(\bbeta)},\notag
\end{align}
which can be seen to be non-concave, also with respect to only the variable block $\bbeta_m$. Thus, following the successive lower-bound maximization, \eqref{Prob:SumRateSub} will be tackled by sequential optimization. To this end, it is necessary to derive a lower-bound of the objective of \eqref{Prob:SumRateSub}, which fulfills Properties \textbf{P1}, \textbf{P2}, and \textbf{P3}, while at the same time leading to a simple optimization problem. To this end, the following lemma proves useful. 
\begin{lemma}\label{Lemma:Conc1}
The function $f:(x,y)\in\mathbb{R}^{2}\to\sqrt{xy}$ is jointly concave in $x$ and $y$, for $x,y>0$.
\end{lemma}
\begin{IEEEproof}
The proof follows upon computing the Hessian of $\sqrt{xy}$ and showing that it is negative semi-definite. Details are omitted due to space constraints.
\end{IEEEproof}
%\begin{IEEEproof}
%We show the result by proving that the Hessian matrix of $f$ is negative semi-definite. After some elaborations, the Hessian of $f$ is equal to
%\begin{equation}
%{\cal H}=\frac{1}{4}\left(\begin{array}{cc}
%-\ds\frac{\sqrt{y}}{x\sqrt{x}} & \ds\frac{1}{\sqrt{yx}} \\\ds\frac{1}{\sqrt{yx}} & -\ds\frac{\sqrt{x}}{y\sqrt{y}}\end{array}\right)
%\end{equation}
%A symmetric matrix is negative semidefinite if all the determinants of its principal sub-matrices are non-positive. This is true for the case at hand, because $-\ds\frac{\sqrt{y}}{x\sqrt{x}}\leq 0$ for $x,y>0$, and the determinant of ${\cal H}$ is equal to zero.
%\end{IEEEproof}
Lemma \ref{Lemma:Conc1}, coupled with the facts that the function $\log_2|(\cdot)|$ is matrix-increasing, and that summation preserves concavity, implies that the rate function in \eqref{Eq:kRate} is the difference of two concave functions. This observation is instrumental for the derivation of the desired lower-bound. Indeed, recalling that any concave function is upper-bounded by its Taylor expansion around any given point $\bbeta_{m,0}$, a concave lower-bound of ${\cal R}_k$ is obtained as
\begin{align}\label{Eq:BoundRate}
{\cal R}_k(\bbeta)&=g_1(\bbeta_m)-g_2(\bbeta_m)\\
&\geq g_1(\bbeta_m)-g_2(\bbeta_{m,0})-\nabla_{\bbeta_m}^Tg_2\rvert_{\bbeta_{m,0}}(\bbeta-\bbeta_{m,0})\notag\\
&=\widetilde{{\cal R}}_k(\bbeta_m,\bbeta_{m,0})\;.\notag
\end{align}
Moreover, it is easy to check that $\widetilde{{\cal R}}_k$ fulfills by construction also properties \textbf{P2} and \textbf{P3} with respect to ${\cal R}_k$. Thus, Problem \eqref{Prob:SumRateSub} can be tackled by the sequential optimization framework, by defining the $\ell$-th problem of the sequence, ${\cal P}_{\ell}$, as the convex optimization program:
\begin{subequations}\label{Prob:SumRateApp}
\begin{align}
&\ds\max_{\bbeta_m}\;\sum_{k=1}^K\widetilde{\mathcal{R}}_{k}(\bbeta_m,\bbeta_{m,0},\bbeta_{-m})\label{Prob:aSumRateApp}\\
&\;\textrm{s.t.}\; \sum_{k\in{\cal K}_m}\eta_{k,m}\leq P_{max,m}\label{Prob:bSumRateApp}\\
&\;\;\;\quad\eta_{k,m}\geq 0\;,\forall\;k\in {\cal K}_m\label{Prob:cSumRateApp}
\end{align}
\end{subequations}
For any $\bbeta_{m,0}$, Problem \eqref{Prob:MinRateApp} can be  solved by means of standard convex optimization theory, since the objective is concave, and the constraints are affine. The resulting power control procedure can be stated as in Algorithm \ref{Alg:SRMax}. Moreover, based on the general theory reviewed in Section \ref{Sec:Optimization}, the following result holds. 
\begin{algorithm}[t]
\caption{Sum rate maximization}
\begin{algorithmic}[1]
\label{Alg:SRMax}
\STATE Set $i=0$ and choose any feasible ${\bbeta}_{2},\ldots,\bbeta_{M}$;
\REPEAT
\FOR{$m=1\to M$}
\REPEAT
\STATE Choose any feasible ${\bbeta}_{m,0}$;
\STATE Let $\bbeta_{m}^{*}$ be the solution of \eqref{Prob:SumRateApp};
\STATE $\bbeta_{m,0}=\bbeta_{m}^{*}$;
\UNTIL{convergence}
\STATE $\bbeta_{m}=\bbeta_{m}^{*}$;
\ENDFOR
\UNTIL{convergence}
\end{algorithmic}
\end{algorithm} 

\begin{proposition}\label{Prop:SRmax}
After each iteration in Line 6 of Algorithm \ref{Alg:SRMax}, the sum-rate value $\sum_{k=1}^K{\cal R}_k$ is not decreased, and the resulting sequence $\{\sum_{k=1}^K{\cal R}_k\}$ converges. Moreover, every limit point of the sequence $\{\bbeta_m\}_m$ fulfills the \gls{kkt} first-order optimality conditions of Problem \eqref{Prob:SumRateSub}. 
\end{proposition}
Two remarks are now in order. First of all an extreme case of Algorithm \ref{Alg:SRMax} is that in which only one variable block is used, namely optimizing all of the transmit powers simultaneously. In this scenario, Algorithm \ref{Alg:SRMax} reduces to a pure instance of sequential optimization, and no alternating optimization is required. Nevertheless, as already mentioned, the complexity of this approach seems prohibitive for large $M$ and $K$.
Then, another extreme case  is that in which the $KM$ transmit powers $\eta_{k,m}$ are optimized one at a time, thus leading to considering $KM$ variable blocks. The advantage of this approach is that each subproblem \eqref{Prob:SumRateApp} would have only one optimization variable, and thus could be solved in closed-form. This brings drastic computational complexity savings and proves to be useful especially in the CF scenario, since in this case each variable block $\bbeta_m$ always has dimension $K$. 
 
\subsection{Minimum rate maximization}
Consider Problem \eqref{Prob:MinRate}. Following similar steps as in Section \ref{Sec:SumRate}, Problem \eqref{Prob:MinRate} with respect to variable block $\bbeta_m$ becomes
\begin{subequations}\label{Prob:SubMinRate}
\begin{align}
&\ds\max_{\bbeta_m}\;\min_{1\leq k\leq K}\; \mathcal{R}_{k}(\bbeta_m,\bbeta_{-m})\label{Prob:aSubMinRate}\\
&\;\textrm{s.t.}\; \sum_{k\in{\cal K}_m}\eta_{k,m}\leq P_{max,m}; 
\quad\eta_{k,m}\geq 0\;,\forall k\in {\cal K}_m\label{Prob:cSubMinRate}
\end{align}
\end{subequations}
Besides the difficulties already encountered in the sum-rate scenario, Problem \eqref{Prob:SubMinRate} poses the additional challenge of having a non-differentiable objective due to the $\min(\cdot)$ operator. To circumvent this issue, \eqref{Prob:SubMinRate} can be equivalently reformulated as the program:
\begin{subequations}\label{Prob:MinRate2}
\begin{align}
&\ds\max_{\bbeta_m,t}\;t\label{Prob:aMinRate2}\\
&\;\textrm{s.t.}\; \sum_{k\in{\cal K}_m}\eta_{k,m}\leq P_{max,m}\label{Prob:bMinRate2}\\
&\;\;\;\quad\eta_{k,m}\geq 0\;,\forall\;k\in {\cal K}_m\label{Prob:cMinRate2}\\
&\;\quad\;\;\mathcal{R}_{k}(\bbeta_m,\bbeta_{-m})\geq t\;,\forall\;k=1,\ldots,K\label{Prob:cMinRate2}\;.
\end{align}
\end{subequations}
At this point, it is possible to tackle \eqref{Prob:MinRate2} by the sequential method. Leveraging again the bound in \eqref{Eq:BoundRate} leads to considering the approximate problem 
\begin{subequations}\label{Prob:MinRateApp}
\begin{align}
&\ds\max_{\bbeta_m,t}\;t\label{Prob:aMinRateApp}\\
&\;\textrm{s.t.}\; \sum_{k\in{\cal K}_m}\eta_{k,m}\leq P_{max,m}\label{Prob:bMinRateApp}\\
&\;\;\;\quad\eta_{k,m}\geq 0\;,\forall\;k\in {\cal K}_m\label{Prob:cMinRateApp}\\
&\;\quad\;\;\widetilde{{\cal R}}_k(\bbeta_m,\bbeta_{m,0},\bbeta_{-m})\geq t\;,\forall\;k=1,\ldots,K\label{Prob:cMinRateApp}\;.
\end{align}
\end{subequations}
For any $\bbeta_{m,0}$, Problem \eqref{Prob:MinRateApp} can be  solved by means of standard convex optimization theory, since the objective is linear, and the constraints are all convex. The resulting power control procedure can be stated as in Algorithm \ref{Alg:MRMax}, which enjoys similar properties as Algorithm \ref{Alg:SRMax}.

\begin{algorithm}[t]
\caption{Minimum rate maximization}
\begin{algorithmic}[1]
\label{Alg:MRMax}
\STATE Set $i=0$ and choose any feasible ${\bbeta}_{2},\ldots,\bbeta_{M}$;
\REPEAT
\FOR{$m=1\to M$}
\REPEAT
\STATE Choose any feasible ${\bbeta}_{m,0}$;
\STATE Let $\bbeta_{m}^{*}$ be the solution of \eqref{Prob:MinRateApp};
\STATE $\bbeta_{m,0}=\bbeta_{m}^{*}$;
\UNTIL{convergence}
\STATE $\bbeta_{m}=\bbeta_{m}^{*}$;
\ENDFOR
\UNTIL{convergence}
\end{algorithmic}
\end{algorithm}

\begin{figure}[t]
\centering
\includegraphics[scale=0.46]{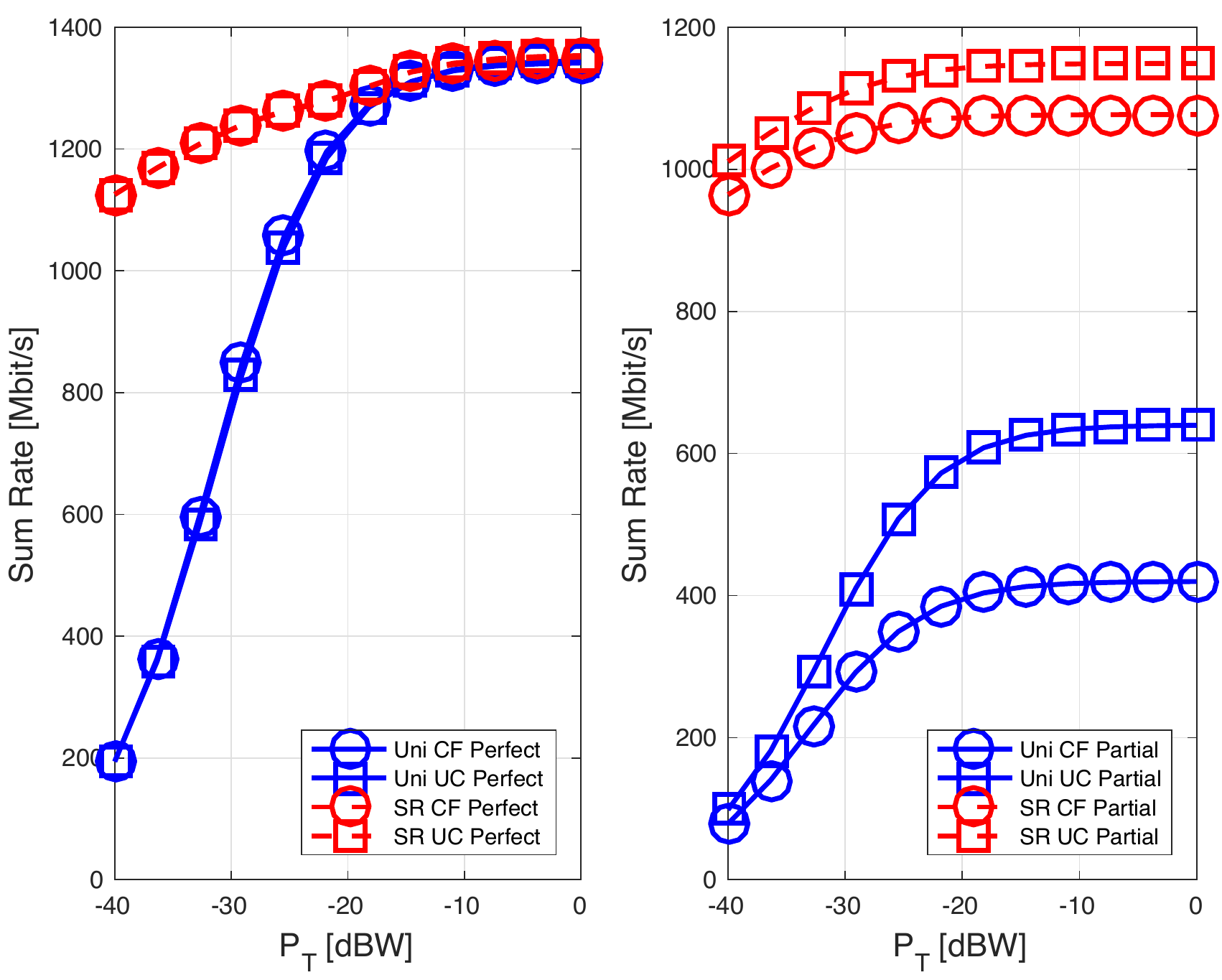}
\caption{Downlink achievable sum-rate for the case of uniform power allocation and of sum-rate maximizing power allocation.}
\label{Fig:SR}
\end{figure}

\begin{figure}[t]
\centering
\includegraphics[scale=0.46]{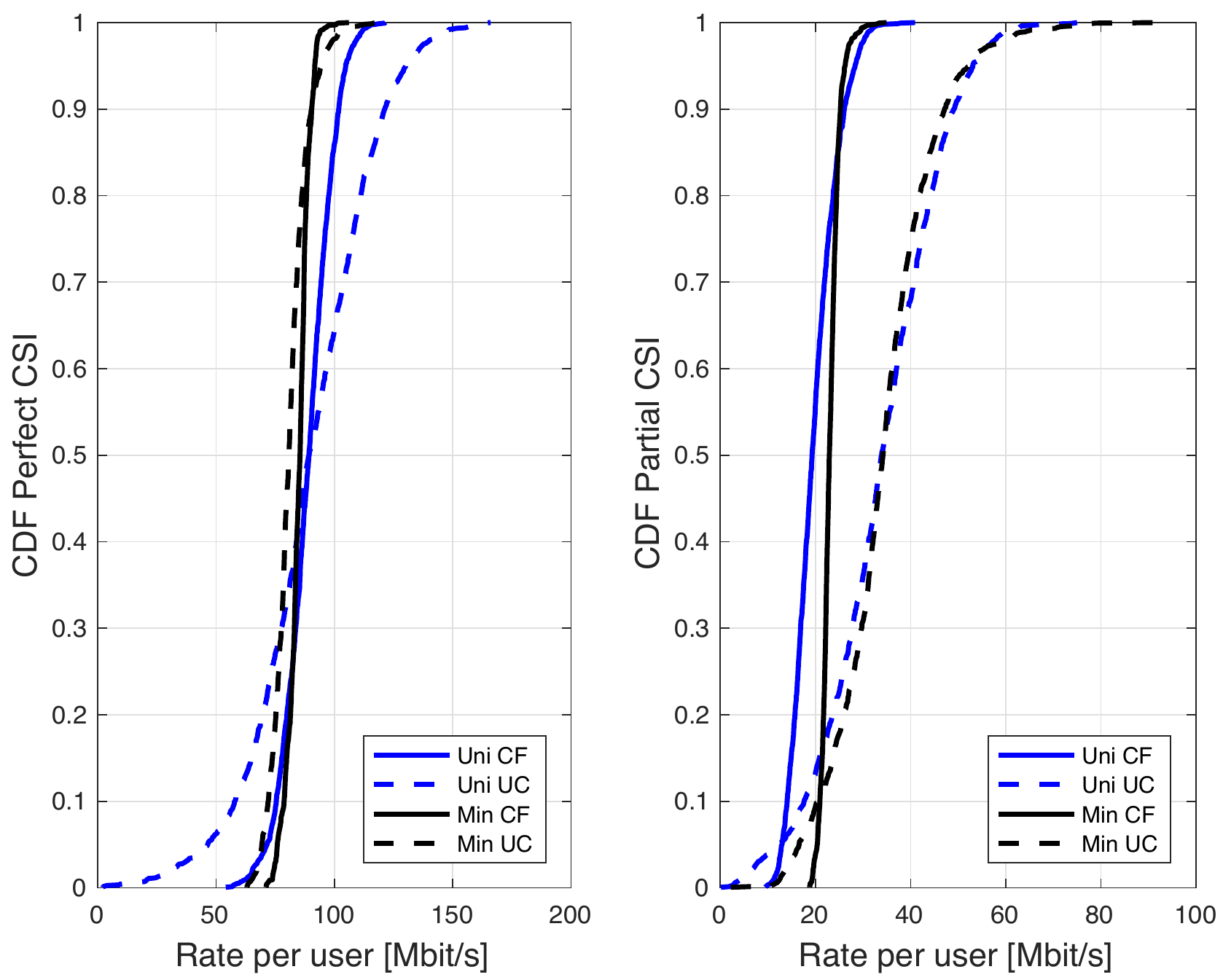}
\caption{Downlink average achievable rate per user CDF for the case of uniform power allocation and of minimum-rate maximizing power allocation.}
\label{Fig:CDF}
\end{figure}
\section{Numerical Results}
In our simulation setup, we consider a communication bandwidth of $W = 20$ MHz centered
over the carrier frequency $f_0=1.9$ GHz. The antenna height at the AP is $15$ m and at the MS is $1.65$ m. The standard deviation of the shadow fading is $\sigma_{\rm sh}=8$ dB, the parameters for the three slope path loss model in \eqref{path_loss} are $d_1=50$ m and $d_0=10$ m, the parameter $\delta$ in \eqref{shadowing} is 0.5 and the correlation distance in \eqref{shadowing_corr} is $d_{\rm decorr}=100$ m. The additive thermal noise is assumed to have a power spectral density of $-174$ dBm/Hz, while the front-end receiver at the AP and at the MS is assumed to have a noise figure of $6$ dB. The shown results come from an average over 100 random scenario realizations with independent MSs and APs locations and channels. 
We quantitatively study and compare the performance of the CF and UC massive MIMO architectures.  We consider $M=60$ APs and $K=15$ MSs spread over a square area of $800 \times 800$ sqm.; we assume $N_{\rm AP}=4$, $N_{\rm MS}=2$ and $P_k=2, \; \forall k=1,\ldots,K$. 
The MSs transmit power is 0.1 W during uplink channel estimation, and the pilot length is $\tau_p=32$. 
Fig. 1 reports the system sum-rate, versus the downlink maximum transmit power per AP $P_{T}$, for the CF and UC approaches, and for the case in which uniform power allocation or sum-rate maximizing power allocation is used. Both the cases of perfect channel state information (CSI) and of partial (i.e., estimated) CSI are considered. Results show that the sum-rate maximizing power allocation provides much better results than the uniform power allocation, especially for the case of partial CSI. Also, while the CF and UC approaches are practically equivalent in the perfect CSI scenario, the UC approach outperforms the CF one in the relevant scenario of partial CSI. Indeed, for partial CSI, APs that are very far from the MS of interest perform very noisy channel estimates, and so their contribution to the communication process is detrimental; this effect is instead not observed in the case of perfect CSI.
Fig. 2 shows the CDF of the rate-per-user for the case of uniform power allocation and minimum rate maximizing power allocation. It is seen that the latter strategy is  effective in increasing the system fairness and in maximizing key performance measures such as the $95 \%$-likely-per-user throughput. As an example, for the case of UC approach with partial CSI, when the proposed power control algorithm is used in place of uniform power allocation, this number increases from 12.6 Mbit/s to 16.5 Mbit/s (+30$\%$).

\section{Conclusion}
Downlink power control algorithms aimed at maximizing either the sum-rate or the minimum rate for a CF massive MIMO system have been introduced, and sample numerical results have been provided. Current work is focused on the consideration of power allocation strategies maximizing the system energy efficiency.

\bibliographystyle{IEEEtran}
%\nocite{*}
\bibliography{FracProg_SB,finalRefs,references}

\end{document}